\documentclass[sigconf]{acmart}

\copyrightyear{2022} 
\acmYear{2022} 
\setcopyright{rightsretained} 
\acmConference[CHI '22]{CHI Conference on Human Factors in Computing Systems}{April 29-May 5, 2022}{New Orleans, LA, USA}
\acmBooktitle{CHI Conference on Human Factors in Computing Systems (CHI '22), April 29-May 5, 2022, New Orleans, LA, USA}
\acmDOI{10.1145/3491102.3502102}
\acmISBN{978-1-4503-9157-3/22/04}

\usepackage{scalerel}
\usepackage{multirow}
\usepackage{enumitem}
\usepackage[utf8]{inputenc}

\newcommand{\system}{\emph{Symphony}}
\newcommand{\feedback}[2][]{\emph{``#2''} #1}
\newcommand{\creator}{ML practitioner}
\newcommand{\consumer}{stakeholder}

\usepackage{xspace,xpunctuate}

\newcommand{\eg}{{e.g.,}\xspace}
\newcommand{\ea}{{et~al.}\xspace}

\newcommand{\etc}{{etc.}\xspace}

\newcommand{\creation}{}
\newcommand{\exploration}{}
\newcommand{\specialization}{}

\begin{document}

\title[Symphony]{Symphony: Composing Interactive Interfaces for Machine Learning}

\author{Alex Bäuerle}
\authornote{Both authors contributed equally to this research.}
\authornote{Work done at Apple.}
\affiliation{%
  \institution{Ulm University}
  \city{Ulm}
  \country{Germany}
}
\email{alex.baeuerle@uni-ulm.de}

\author{Ángel Alexander Cabrera}
\authornotemark[1]
\authornotemark[2]
\affiliation{%
  \institution{Carnegie Mellon University}
  \city{Pittsburgh}
  \state{PA}
  \country{USA}
}
\email{cabrera@cmu.edu}

\author{Fred Hohman}
\affiliation{%
  \institution{Apple}
  \city{Seattle}
  \state{WA}
  \country{USA}
}
\email{fredhohman@apple.com}

\author{Megan Maher}
\affiliation{%
  \institution{Apple}
    \city{Cupertino}
  \state{CA}
  \country{USA}
}
\email{megan_maher@apple.com}

\author{David Koski}
\affiliation{%
  \institution{Apple}
      \city{Cupertino}
  \state{CA}
  \country{USA}
}
\email{dkoski@apple.com}

\author{Xavier Suau}
\affiliation{%
  \institution{Apple}
  \city{Barcelona}
  \country{Spain}
}
\email{xsuaucuadros@apple.com}

\author{Titus Barik}
\affiliation{%
  \institution{Apple}
  \city{Seattle}
  \state{WA}
  \country{USA}
}
\email{tbarik@apple.com}

\author{Dominik Moritz}
\affiliation{%
  \institution{Apple}
  \city{Pittsburgh}
  \state{PA}
  \country{USA}
}
\email{domoritz@apple.com}

\renewcommand{\shortauthors}{Bäuerle and Cabrera, \ea}

\begin{abstract}
Interfaces for machine learning (ML), information and visualizations about models or data, can help practitioners build robust and responsible ML systems.
Despite their benefits, recent studies of ML teams and our interviews with practitioners (n=9) showed that ML interfaces have limited adoption in practice.
While existing ML interfaces are effective for specific tasks, they are not designed to be reused, explored, and shared by multiple stakeholders in cross-functional teams.
To enable analysis and communication between different \creator{}s, we designed and implemented \system{}, a framework for composing interactive ML interfaces with task-specific, data-driven components that can be used across platforms such as computational notebooks and web dashboards.
We developed \system{} through participatory design sessions with 10 teams (n=31), and discuss our findings from deploying \system{} to 3 production ML projects at Apple.
\system{} helped \creator{}s discover previously unknown issues like data duplicates and blind spots in models while enabling them to share insights with other stakeholders.
\end{abstract}



\begin{teaserfigure}
    \centering
    \includegraphics[width=0.95\textwidth]{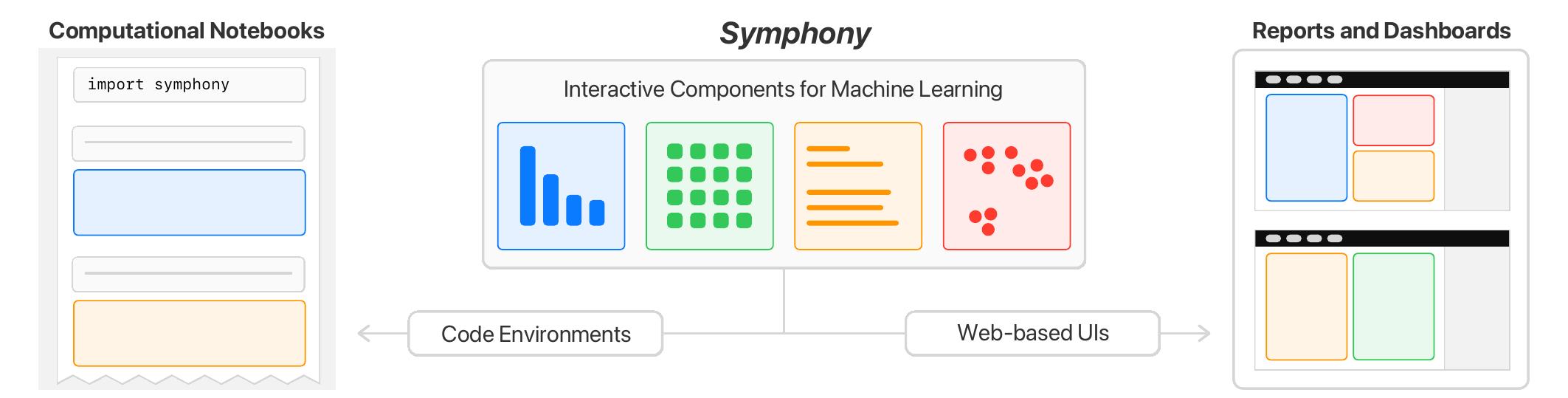}
    \caption{
    \system{} applies techniques from machine learning (ML) documentation, data visualization, and interactive programming to create ML interfaces with interactive, task-specific components.
    Diverse \creator{}s can explore their data and analyze their models where they work, both in computational notebooks and in web-based dashboards.
    }
    \Description{A diagrammatic figure showing the workflow of Symphony. In the middle is a collection of Symphony components, with an arrow going to the left to a Jupyter notebook and an arrow going to the right for a standalone dashboard.}
    \label{fig:teaser}
\end{teaserfigure}

\maketitle

\section{Introduction}

Successfully deploying machine learning systems in production is a complex, collaborative process that involves a wide range of~\creator{}s, from data scientists and engineers to domain experts and product managers. 
A substantial amount of research has gone into creating ML interfaces for analyzing and sharing insights about ML systems that practitioners can use to better understand and improve deployed ML products.
We describe \textit{machine learning interfaces} as static or interactive artifacts, visualizations, and information that communicate details about ML data and models.
ML interfaces include documentation methods (\eg Model Cards~\cite{mitchell2019model}, Datasheets~\cite{gebru2018datasheets}), visualization dashboards (\eg What-if Tool~\cite{Wexler2019}, ActiVis~\cite{Kahng2018}, among many others~\cite{hohman2018visual}), and interactive programming widgets (\eg ipywidgets~\cite{ipywidgets}, Streamlit~\cite{Streamlit}) that give practitioners insights into what their datasets contain and how their models behave.
Despite the benefits and breadth of ML interfaces, recent studies have found that they are not as widely used and shared in practice as expected~\cite{zhang2020, Koesten2019}.
This underuse can lead to missed data errors and model failures, a lack of shared team understanding of model behavior, and, ultimately, deployed ML systems that may be biased~\cite{buolamwini2018gender} or unsafe~\cite{Oakden-Rayner2020}.

To understand why ML interfaces are not used more frequently, we interviewed 9 \creator{}s at Apple about their current machine learning practice and workflows.
We found that while \creator{}s want to use them, current interfaces have limitations that make them either insufficient or too time consuming to use.
One category of ML interfaces are \textit{ML documentation} methods, such as Model Cards~\cite{mitchell2019model} and Datasheets~\cite{gebru2018datasheets}, which describe the details and records the provenance of an ML system's data and model.
Documentation methods often lack the interactive tools and visualizations necessary for specific analyses and have to be manually authored and updated separately from where ML development happens.
Another category of interfaces, \textit{visualization dashboards}, consist of multiple coordinated views tailored to specific domains and tasks.
\creator{}s must learn a new platform and wrangle their data into the right format in order to use these bespoke systems, which also require significant work to reuse for different tasks.
Finally, \textit{interactive programming widgets} can render web-based ML visualizations directly in code environments.
However, widgets typically cannot be used outside of the platform in which they were created and often lack complex visualizations required by modern ML models and unstructured data---non-tabular data types such as images, videos, audio, point-clouds, sensor data, \etc
Overall, we found that while current ML interfaces work well for specific tasks and platforms, they are not designed to be reused, explored, and shared by diverse stakeholders in cross-functional ML teams.

\begin{figure*}
    \centering
    \includegraphics[width=\textwidth]{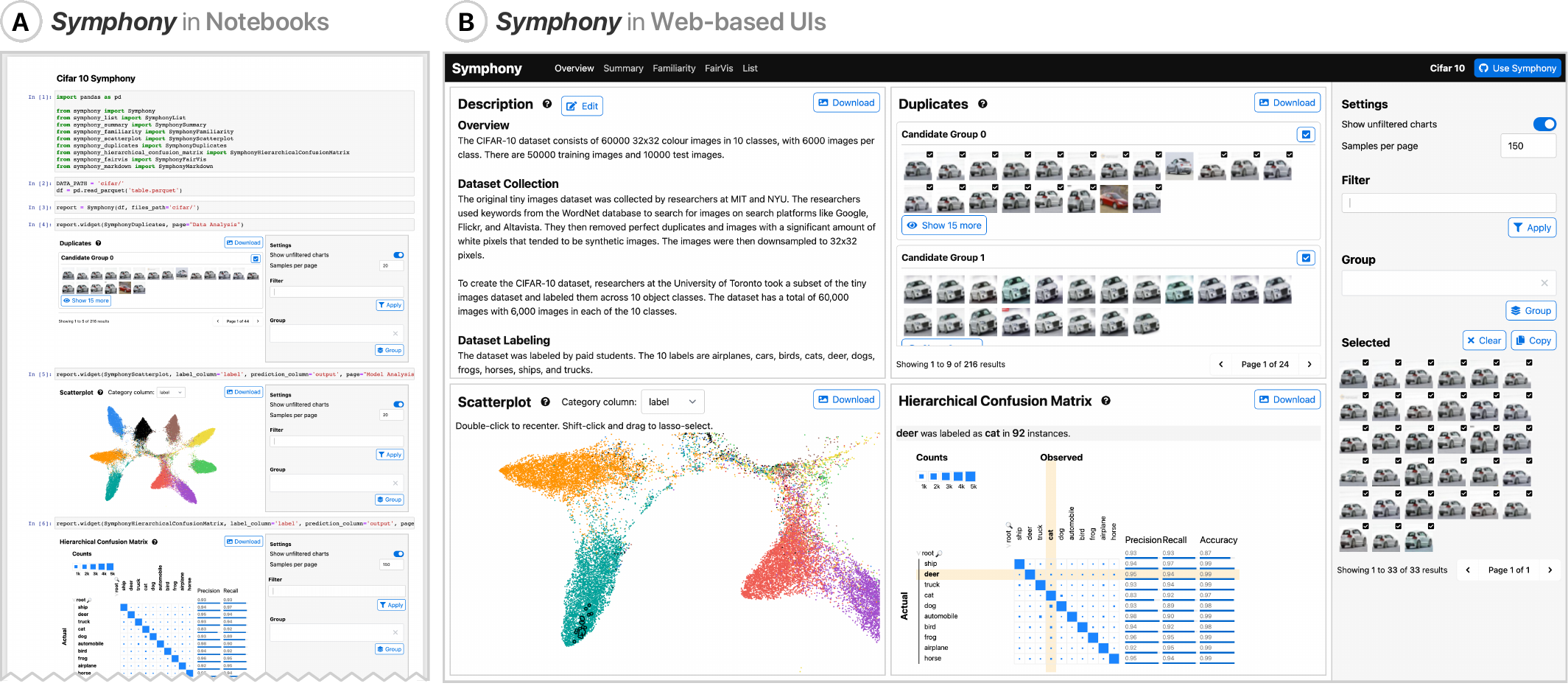}
    \caption{
    A demonstration of \system{} running in both (A) a computational notebook and (B) a web-based UI with the same visualization components and code.
    In a computational notebook, an \creator{} passes their data and model outputs directly from Python variables like Pandas Data Frames~\cite{mckinney-proc-scipy-2010} to \system{} components.
    The \creator{} can then export the components to a self-contained, web-based UI.
    This example shows \system{} loaded with the CIFAR-10~\cite{krizhevsky2009learning} dataset and a trained image classification model.
    After reading a textual description of the dataset, a user found and selected duplicate car instances which were reactively highlighted in the projection component and the confusion matrix.
    The user then explored the confusion matrix to determine if the duplicates could be impacting model performance.
    }
    \Description{Two side by side screenshots, one of a Jupyter notebook with Symphony components and the same components in a UI dashboard. They each have four components, a markdown description, a list of duplicates, an embedding projection, and a confusion matrix.}
    \label{fig:notebook-and-overview}
\end{figure*}

Our formative research showed that ML work requires bespoke visualizations for complex models and data types which work across the different platforms \creator{}s use.
To address these needs, we combined the affordances of existing ML interfaces to design and implement \textbf{\system{}}, a framework for creating and composing interactive ML interfaces with task-specific, data-driven visualization components.
\system{} supports two popular platforms used by \creator{}s, code environments such as Jupyter notebooks and no-code environments such as web-based UIs (Figure~\ref{fig:teaser}).
\system{} components are JavaScript modules that use custom code or existing libraries to create task-specific visualizations of structured and unstructured data.
Each component is also fully interactive: users can filter, group, or select instances either through a UI toolbar or code.
These interactions are reactively synchronized across \system{} components, enabling linked visualizations.
\system{}'s cross-platform availability enables \creator{}s to use the same components for both exploring \textit{and} sharing insights about their ML systems (Figure~\ref{fig:notebook-and-overview}).

We worked with ML teams at Apple to both design \system{} and apply it to deployed ML projects.
To collect the diverse requirements and use cases for ML interfaces, we conducted participatory design sessions with 10 ML teams with a total of 31 \creator{}s.
Informed by these sessions, we implemented a set of 11 components supporting a range of different models and data types.
We then worked with 3 teams from the design sessions to deploy \system{} in their machine learning workflows and ran a think-aloud study with them to qualitatively evaluate \system{}.

Teams using \system{} with their real-world data and models found surprising insights which they had not previously known, such as duplicate instances, labeling errors, and model blind spots.
Participants also described a variety of use cases for \system{}, from creating automated dataset reports to analyzing model performance in computational notebooks.
Moreover, participants that did not previously share their analyses also showed interest in using \system{} in their teams to better communicate the state of their ML system with other stakeholders.

The main contribution of this work is \system{}, a framework for composing interactive ML interfaces with task-specific, data-driven visualization components.
To design \system{}, we conducted formative interviews, participatory design sessions, and case studies on deployed ML workflows with a total of 39 \creator{}s across 15 teams.
\system{} enabled \creator{}s to discover significant issues like dataset duplicates and model blind spots, and encouraged them to share their insights with other stakeholders.
\system{} combines the following principles to improve upon existing ML interfaces:

\begin{itemize}
    \item \textbf{Data-driven ML interfaces} derived from and updated with ML data and models.
    \item \textbf{Task-specific visualizations} for unstructured data and modern machine learning models.
    \item \textbf{Interactive exploration tools} for exploring different dimensions of an ML system.
    \item \textbf{Reusable components} that can be used, composed, and shared across different platforms.
\end{itemize}

\section{Background and Related Work}

\system{} bridges three areas of related work: ML documentation methods, data visualization dashboards, and interactive programming environments.
First, the \system{} framework can be used to write and share ML documentation.
Second, \system{} components can show complex visualizations and be composed into visual analytics dashboards to help \creator{}s make sense of ML data and models (Section~\ref{sec:vast}).
Lastly, \system{} components can be used in and exported from interactive programming environments, like computational notebooks, which are often used by \creator{}s (Section~\ref{sec:notebooks}).

\subsection{Documenting Data and Models}\label{sec:reports}
A variety of documentation methods exist to help \creator{}s track and communicate details about their data and models.
Without knowing what a dataset contains or what a model has learned, teams can inadvertently release AI products with issues like safety concerns and biases~\cite{shneiderman2020bridging,floridi2019establishing,aiguidelines,silberg2019notes}, as seen in numerous deployed systems~\cite{buolamwini2018gender,snow2018amazon,wilson2019predictive,hendricks2018women}.

Since machine learning models are a direct result of the data they were trained on, it is important to first understand the data behind an ML system.
Datasheets for Datasets~\cite{gebru2018datasheets} applies the idea of datasheets in electrical engineering to describe important attributes of a dataset, such as collection methods and intended uses.
Similar work has focused on specific types of data, for example, Data Statements~\cite{bender2019data} are tailored to natural language processing datasets.
These guidelines describe \textit{what} should be included in documentation, not \textit{how} an author can create or share the resulting artifact~\cite{Koesten2019}.
Additionally, these documents are static \LaTeX{} or text documents that are disjoint from the backing data and models and have to be manually updated.
Since there is heterogeneity in what information is important for each dataset, \citet{holland2018dataset} proposed the more general concept of Dataset Nutrition Labels, modular graphs describing different aspects of a dataset. 
Like \system{}, these labels use modular visualizations, however, they focus on simple aggregate visualizations without displaying data samples and do not support platforms where \creator{}s do their work.

A parallel line of research has focused on documenting machine learning models.
Model Cards~\cite{mitchell2019model} and FactSheets~\cite{arnold2019factsheets} are similar concepts to Datasheets that can include important information and details about machine learning models.
These model reports include information ranging from the model type and hyperparameters to aggregate metrics and ethical considerations.
Similar to Datasheets, these types of documentation are disjoint from the backing data and do not include interactive visualizations of model details and performance metrics.

\subsection{Visualization for Machine Learning} \label{sec:vast}
There are a growing number of visualization systems that help \creator{}s make sense of modern ML systems with unstructured datasets and machine learning models~\cite{hohman2018visual}.
Visualizations can help \creator{}s in tasks such as auditing models for bias~\cite{Cabrera2019}, understanding the internals of deep learning architectures \cite{hohman2019summit}, and guiding automatic model selection \cite{Cashman2019}.
A full review of this literature is out of scope for this work, but we provide a sample of representative systems to highlight the types of visualizations that could be implemented as \system{} components.

Data science work often starts with and leads back to understanding the backing data. 
Modern machine learning models and tasks use \textit{unstructured} data like images and audio that cannot be visualized and explored with tables and histograms.
Systems like Know Your Data~\cite{kyd} and Facets~\cite{facets} are visualization dashboards for exploring unstructured data.
Other visual analytics systems process the data further to derive insights like outliers~\cite{Chen2020}, biases in a dataset ~\cite{wang2020revise}, or mislabeled data instances~\cite{xiang2019}.
With a deeper understanding of their data, \creator{}s can more effectively debug and improve their models.

The models \creator{}s use are often large, complex black-box models like deep learning systems.
Visualization systems like Summit~\cite{hohman2019summit} and Seq2Seq-Vis~\cite{Strobelt2019} can help \creator{}s develop a better mental model of how their machine learning systems work and what they are learning.
Another set of systems, including Model Tracker~\cite{Amershi2015}, Squares~\cite{Ren2017}, AnchorViz~\cite{Chen2018}, ConfusionFlow~\cite{Hinterreiter2020}, What-if Tool~\cite{Wexler2019}, and MLCube~\cite{Kahng2016}, focus on performance analysis and provide different views of a model's errors 
Lastly, there are tools for detecting potential biases~\cite{Ahn2019} or systematic errors~\cite{Wu2019,bauerle2020classifier} in training data.
These various of visualizations can be repackaged as \system{} components, for example, we implement a version of FairVis~\cite{Cabrera2019} as a component for auditing classifiers for bias.

Lastly, there are integrated systems that help \creator{}s both implement and visualize ML models.
One of the first systems describing such an integrated system is Gestalt~\cite{patel_gestalt_2010}, a development environment with visualizations for training and analyzing classification models.
A subsequent system focused on interactive machine learning is Marcelle~\cite{francoise_marcelle_2021}, which uses composable stages and visualizations to create interactive ML interfaces.
In contrast to Gestalt and Marcell, \system{} is focused on the analysis stage of ML systems, and includes important features such as cross-platform support, reactivity, and a consistent data API which are not available in Gestalt and Marcelle.

ML data and model visualizations are often deployed as visual analytics dashboards that are separate from both interactive programming environments that \creator{}s work with and ML documentation shared with other \consumer{}s.
This separation limits who can use visualizations to understand ML data and models.
\system{} aims to bridge these worlds by bringing visualizations both into notebooks where data work happens and into the documentation shared with other \consumer{}s.

\subsection{Interactive Programming Environments}\label{sec:notebooks}

\creator{}s often use interactive programming environments for exploring and modeling data since they can interact with and iterate on their ML systems~\cite{Kery2018}.
These environments are most commonly implemented as computational notebooks like Jupyter \cite{Kluyver2016jupyter}, DataBricks \cite{DataBricks}, and Observable \cite{Observable}.
While computational notebooks have extensions for creating interactive visualizations, such as the ipywidgets API~\cite{ipywidgets} for Jupyter, they are often underused~\cite{Batch2018Gap} and hard to share~\cite{Head2019Messes, Kery2018}.

Several libraries exist for interactively visualizing data in notebooks.
Graphing libraries such as Altair~\cite{VanderPlas2018} and Plotly~\cite{plotly} allow users to create interactive charts but only support a finite set of graphs and require users to manually define what visualizations they want to use.
Lux~\cite{Lee2021Lux} and B2~\cite{Wu2020B2} lower the cost of using visualizations in notebooks by automatically providing relevant charts for users' data frames.
These approaches help analyze tabular data, but they lack the specific visual representations needed for machine learning development.

A separate challenge is sharing visualizations and other notebook outputs outside of the notebook context.
Voilà~\cite{voila} tackles this challenge directly by exporting full Jupyter notebooks to a hosted website.
\creator{}s can use Voilà to share notebooks that contain \system{} components, but it requires a Python kernel to be running and Voilà does not provide any visualizations itself.
Two visualization frameworks similar to \system{}, Panel~\cite{panel} and Plotly Dash~\cite{dash}, use independent components to create visualizations that can be used in both Jupyter notebooks and standalone websites.
However, these tools also have limitations for creating complete ML interfaces: Panel visualizations are tied to the Jupyter ecosystem and lack interactivity without a Python backend, while Plotly Dash primarily supports Plotly charts and does not easily extend to custom visualizations.
\system{} provides components that are fully interactive in both notebooks and web UIs, and support any JavaScript-based visualization.
Additionally, \system{}'s shared state synchronizes its components, enabling reactive brushing and linking between views.

More recent interactive programming environments have moved away from the notebook paradigm.
For example, in the Streamlit~\cite{Streamlit} platform, users write Python scripts using a library that renders interactive components in a separate website.
While Streamlit supports interactive components like Jupyter notebooks, it is primarily an environment focused on designing web applications rather than exploratory data science or ML reporting.
Exploratory analysis is still often done in notebooks, and Streamlit requires users to learn a new platform.
Other platforms are moving away from programming altogether, such as Glinda~\cite{DeLine2021a}, a declarative language that lets \creator{}s describe analysis steps in a domain-specific language.
Glinda does not define any specific visualizations, but it could be complemented by \system{} components.
Since \system{} components are standalone JavaScript modules, future wrappers could integrate \system{} components into data science environments like Streamlit and Glinda.

\section{Formative Interviews}

To understand how ML interfaces are used in practice, we conducted 7 semi-structured interviews with 9 participants at Apple.
We recruited participants through internal emails and messaging boards and selected participants across a range of different roles, including engineers, designers, researchers, and testing roles that work on teams to build and deploy ML systems.
Each interview was conducted over a video call and lasted about an hour.
First, we asked participants about how they currently create and use different ML interfaces like documentation, visualization dashboards, and widgets.
We then asked them what the main limitations and pain points are in current tools and what types of improvements they would find helpful.
From these need-finding interviews we identified the following themes.

\paragraph{Use cases for ML interfaces}
All participants agreed that creating and sharing ML interfaces can help them build more robust and capable ML products.
Participants described use cases of ML interfaces in myriad tasks, such as \feedback[(P2)]{flagging failures for review,} \feedback[(P4)]{detecting systematic failures,} and \feedback[(P1)]{fairness and bias education.}
Participants also mentioned stages across the entire ML process in which ML interfaces can be useful, from \feedback[(P5)]{dataset curation and sharing} to analysis \feedback[(P7)]{after an ML model has been trained,} or \feedback[(P1)]{in all stages}.
Consequently, since different stakeholders involved in an ML product need specific views of the data and models, ML interfaces must be flexible enough to support analysis across numerous tasks and domains.

\paragraph{Ad-hoc tools and analyses}
While all participants detailed clear use cases for ML interfaces, they also mentioned limitations preventing them from using existing tools or sharing insights.
One participant bluntly stated \feedback[(P3)]{right now, we basically have no tools} for analyzing ML systems.
Instead, participants rely on ad-hoc, hand-crafted visualizations for their specific analyses.
For example, one of our participants said their process for looking at instances is to \feedback[(P9)]{manually examine icons in a file explorer.}
Another participant \feedback[(P4)]{looks at handcrafted summaries of select data subsets} to do model analysis.
Larger teams with more resources may have bespoke tools, such as one participant that \feedback[(P6)]{use[s] a team-internal tool to analyze data}.
Overall, a lack of adequate tooling leads to \creator{}s using one-off, manual tools or ML teams investing in their own, custom visualization systems.

\paragraph{Limitations of existing ML interfaces}
Participants detailed a variety of technical roadblocks and time-consuming processes preventing them from using existing ML interfaces.
Many tools require users to wrangle and export their data into a specific format before loading it into a custom system or dashboard.
However, as one participant stated, \feedback[(P1)]{we do not have a lot of time for creating such visualizations:} \creator{}s simply do not have the bandwidth to do the setup and data wrangling work necessary to use separate systems.
\creator{}s' main priority is working on data and models, and \feedback[(P6)]{if it takes longer than 5-10 minutes, I am not going to [use an ML interface] immediately}.

Five participants mentioned explicitly that they do not use ML interfaces because they are not available in the environments where they work, and that \feedback[(P3)]{people would want to use easier tools.}
For example, \feedback[(P2)]{many data scientists want to explore their data in notebooks} without having to open a separate system.
Additionally, since data and models update frequently, one participant wanted to \feedback[(P6)]{start a job with checkboxes and buttons} and produce a self-updating web UI that they would not have to manually author.

Lastly, the teams we talked to work with myriad data types, such as video, 3D point cloud, tabular, image, and audio data, and desired bespoke visualizations supporting their analysis needs.
One participant mentioned running and visualizing specific data analyses, and \feedback[(P8)]{would want to specify algorithms because our problems are very specialized.}
However, current data science tools often only provide visualizations for a limited set of data types and models.

\paragraph{Lack of communication between \consumer{}s}
As a consequence of limited, isolated interfaces, participants described various challenges for communicating and sharing insights.
Since different stakeholders prefer different environments, such as code-based notebooks or standalone dashboards, it can be challenging to share insights with others.
In addition to sharable interfaces, participants also wanted cross-platform support for themselves, as one participant put it, \feedback[(P2)]{I would like both an environment for experimentation and always there reliable visualizations.}

It can also be difficult to transfer visualizations and findings between platforms that different stakeholders work with.
One participant lamented that \feedback[(P4)]{I am often not invited to the table until things go wrong,} and in some teams \feedback[(P3)]{designers often times don't have access to data and model results.}
In turn, decisions about ML systems are made without all team members having a shared understanding of the current state and limitations of the project.
Despite these current limitations, participants thought that \feedback[(P3)]{fostering a culture of sharing insights would be great.}

\section{Design Goals}\label{sec:goalschallenges}

Based on the challenges we identified in the formative interviews, we found that a successful framework for ML interfaces must fulfill the following:

\begin{enumerate}
    \setlength\itemsep{0.5em}\leftskip-24.5pt
    \item[\creation{}] \noindent\textbf{Enable data-driven ML interfaces.}
    ML interfaces are often disconnected from an ML system's backing data and model outputs~\cite{gebru2018datasheets,mitchell2019model}.
    \creator{}s should be able to create visualizations that are up-to-date and reflect an ML systems' current state.
    
    \item[\specialization{}] \noindent\textbf{Support task-specific visualizations.}
    Specialized visualizations are often needed to make sense of the unstructured data and deep learning models increasingly used in machine learning~\cite{vellido2020importance,sacha2017you}.
    ML interfaces should support these task-specific visualization needs.
    
    \item[\exploration{}] \noindent\textbf{Provide interactive exploration tools.}
    Static ML interfaces only show a fixed subset of the possible analyses stakeholders may need~\cite{Koesten2019}.
    Interactive visualizations let different \consumer{}s discover and validate the patterns most relevant to their goals.
    
    \item[] \noindent\textbf{Make components reusable.}
    Different stakeholders explore ML systems in different environments, such as computational notebooks and web-based UIs.
    ML interfaces should be available across environments and reusable for different domains and tasks.
\end{enumerate}

\begin{figure*}[t]
  \centering
  \includegraphics[width=\linewidth]{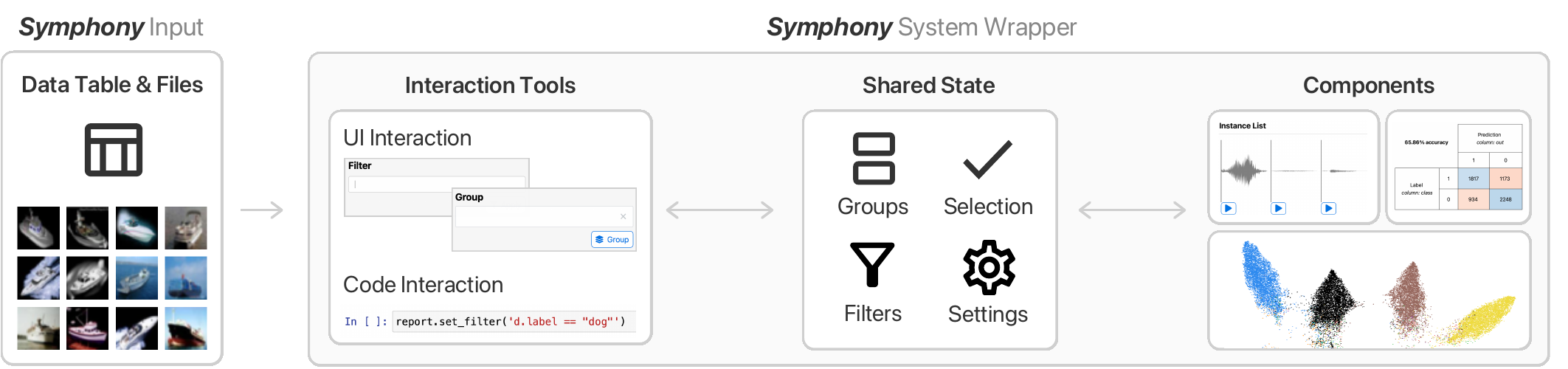}
  \caption{
    The technical overview of the \system{} framework.
    A dataset and files are passed into the \system{} wrapper for a particular platform.
    The wrapper holds the shared state which is reactively updated and modified either by standardized interaction tools or components themselves.
  }
  \Description{A diagrammatic figure showing how the Symphony framework works. On the left is a data table and images that are shown going into a Symphony wrapper that contains interaction tools, shared state, and components.}
  \label{fig:framework_overview}
\end{figure*}

\section{Symphony: A Framework for Composing Interactive Interfaces for Machine Learning}

Based on these design goals we built \system{}, a framework for composing ML interfaces from interactive visualization components.
\creator{}s can explore their data and models using \system{} components in a computational notebook and then combine and transform them into web-based UIs.
\system{} consists of three primary features: modular components (Section~\ref{sec:components}), environment wrappers (Section~\ref{sec:wrappers}), and interaction tools (Section~\ref{sec:exploration}).
In the following, we describe the specific design and implementation choices we made to support these goals.

\subsection{Modular Components}\label{sec:components}

The building blocks of \system{} are independent, modular components designed for task-specific visualizations (Figure~\ref{fig:framework_overview}, right).
A \system{} component is a JavaScript module that renders a web-based visualization. 
We use the Svelte\footnote{https://svelte.dev} web framework as the base of \system{} components, but visualizations can be written using any JavaScript code or library.
JavaScript has a rich ecosystem of libraries and APIs for creating interactive visualizations, like D3 and Three.JS, which can be used to create \system{} components.
This flexibility is important for visualizing unstructured ML datasets, something that is not supported by common charting libraries like Matplotlib~\cite{Hunter:2007} or Altair~\cite{VanderPlas2018}.

Each \system{} component is passed three parameters: a metadata table, derived state variables like grouped tables, and references to raw data instances like images.
The metadata table contains a row for each instance from which a set of \textit{state} variables, such as filtered and grouped tables are derived (state variables are described in detail in Section~\ref{sec:exploration}).
Components are also passed a URL from which to fetch raw data samples such as images or audio files. 
\system{} controls these three parameters, synchronizing and reactively updating them across components.

New components can be created using a cookiecutter template that generates all the boilerplate code needed to integrate components with \system{}.
In the cookiecutter code, a component developer modifies the front-end JavaScript to create their custom interactive visualization.
They can make use of the parameters provided by \system{} to base their visualization on the data provided by a \creator{}.
In the following Subsection we show how these modular, reactive components can then be composed by a \system{} wrapper to be used across different platforms.

\subsection{Platform Wrappers}\label{sec:wrappers}

The primary goal of using self-contained components is to compose and share them as flexible interfaces across different platforms.
This is done using \system{}'s next main feature, wrappers, which connect components with a particular backing platform.
These wrappers have two primary functions - first, passing data from a platform to \system{} in the correct format, and second, rendering \system{} components in the platform's UI.
To support both exploring and sharing ML interfaces, we implemented wrappers for the two platforms most requested in our formative study, Jupyter notebooks and web UIs. 
These platforms are also representative of the two environments we found to be most used by \creator{}s: programming environments for exploratory analysis and web-based UI interfaces for sharing insights.

The Python wrapper bundles \system{} components as packages which can be published to a package index like PyPI for use in notebooks and Python scripts.
To make \system{} interfaces available in Jupyter notebooks, \system{}'s Python wrapper also makes each component an ipywidget~\cite{ipywidgets}.
The ipywidgets API renders web-based widgets in the Jupyter notebook UI and synchronizes its variables with the Python kernel.
Data tables like Pandas DataFrames or Apache Arrow tables, along with an endpoint for raw instance files, can be passed to \system{}'s Python wrapper to connect components to the data.

\vspace*{5pt}
\noindent\includegraphics[width=\linewidth]{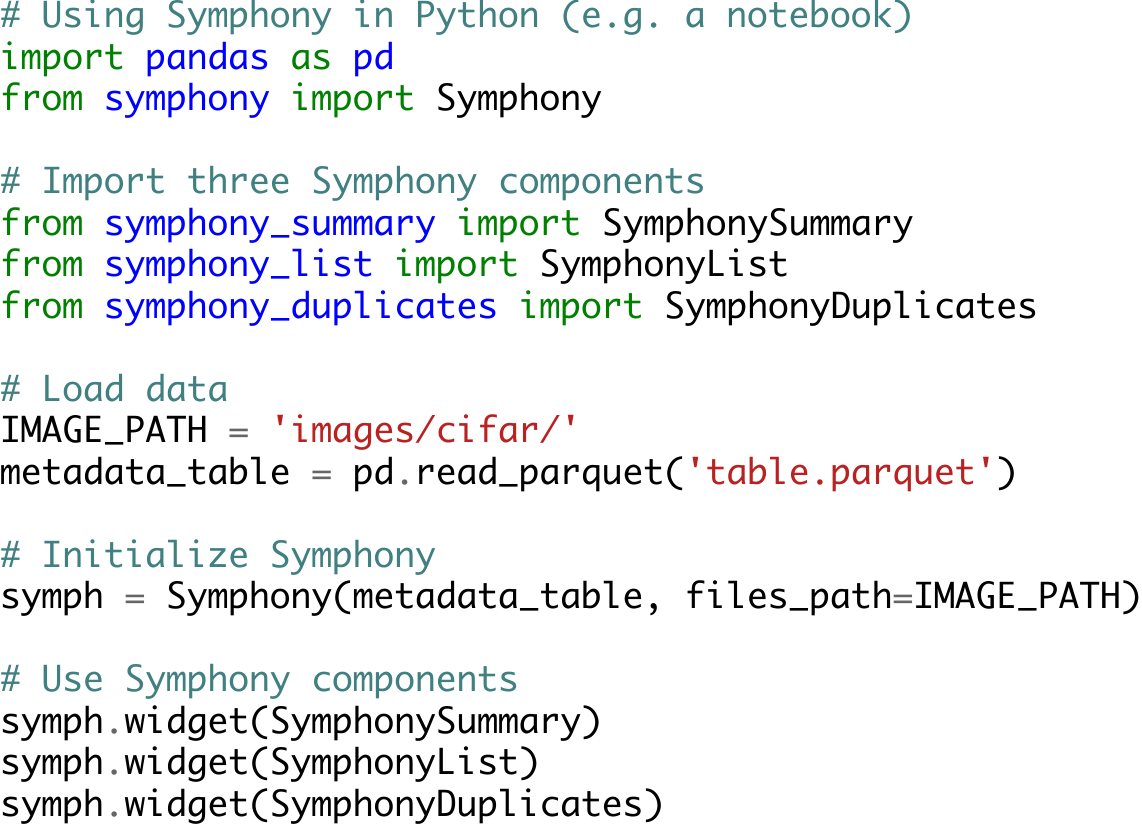}

\begin{figure*}
    \centering
    \includegraphics[width=\textwidth]{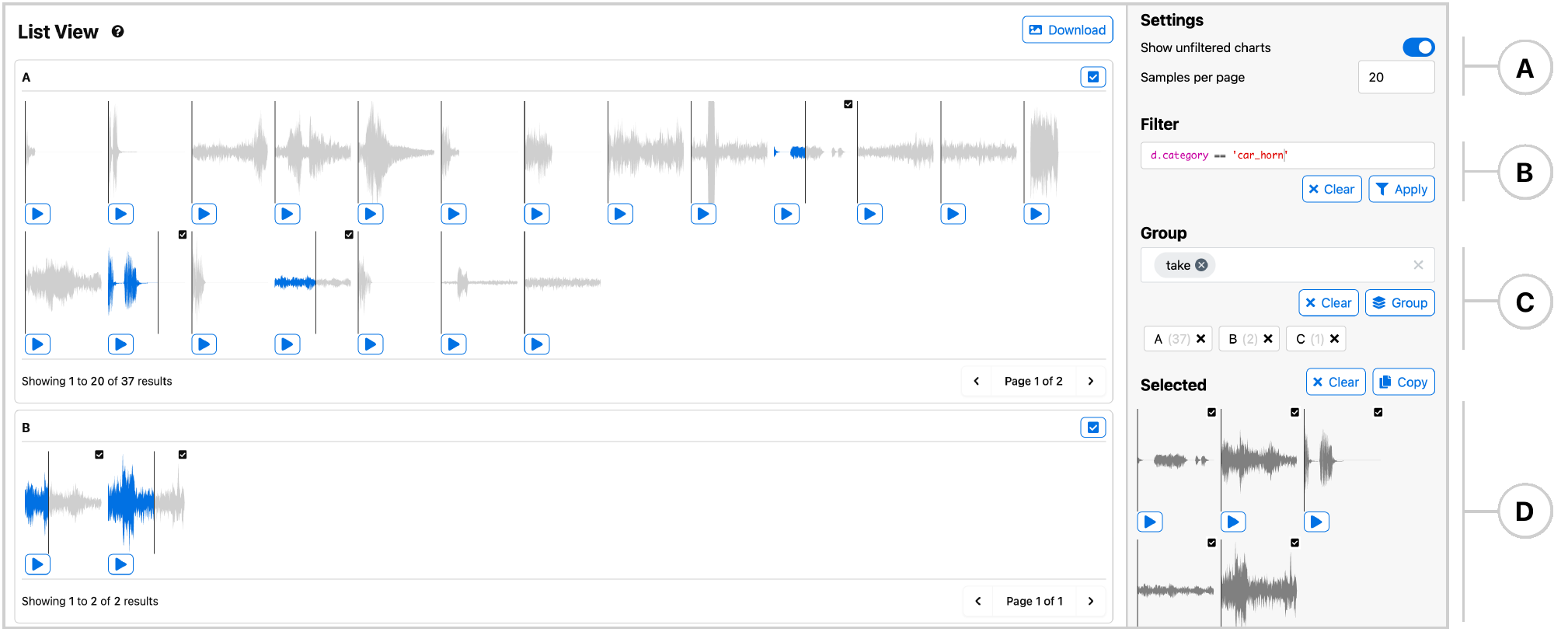}
    \caption{A list component looking at audio samples from the ESC-50 environmental noise classification dataset.
    The toolbar on the right has UI elements for the different interactions tools available in \system{}. 
    The user (A) has increased the number of instances shown per page, and then (B) filtered to see only car horn noises.
    They then (C) grouped by the ``take'' feature, and (D) selected a set of interesting instances.
    In a notebook a user can also set these parameters from code.}
    \Description{A screenshot of an individual component from Symphony. It shows a list of audio waveforms for audio files, and a toolbar on the right hand side with a series of UI settings.}
    \label{fig:component}
\end{figure*}

\begin{figure}
    \centering
    \includegraphics[width=\linewidth]{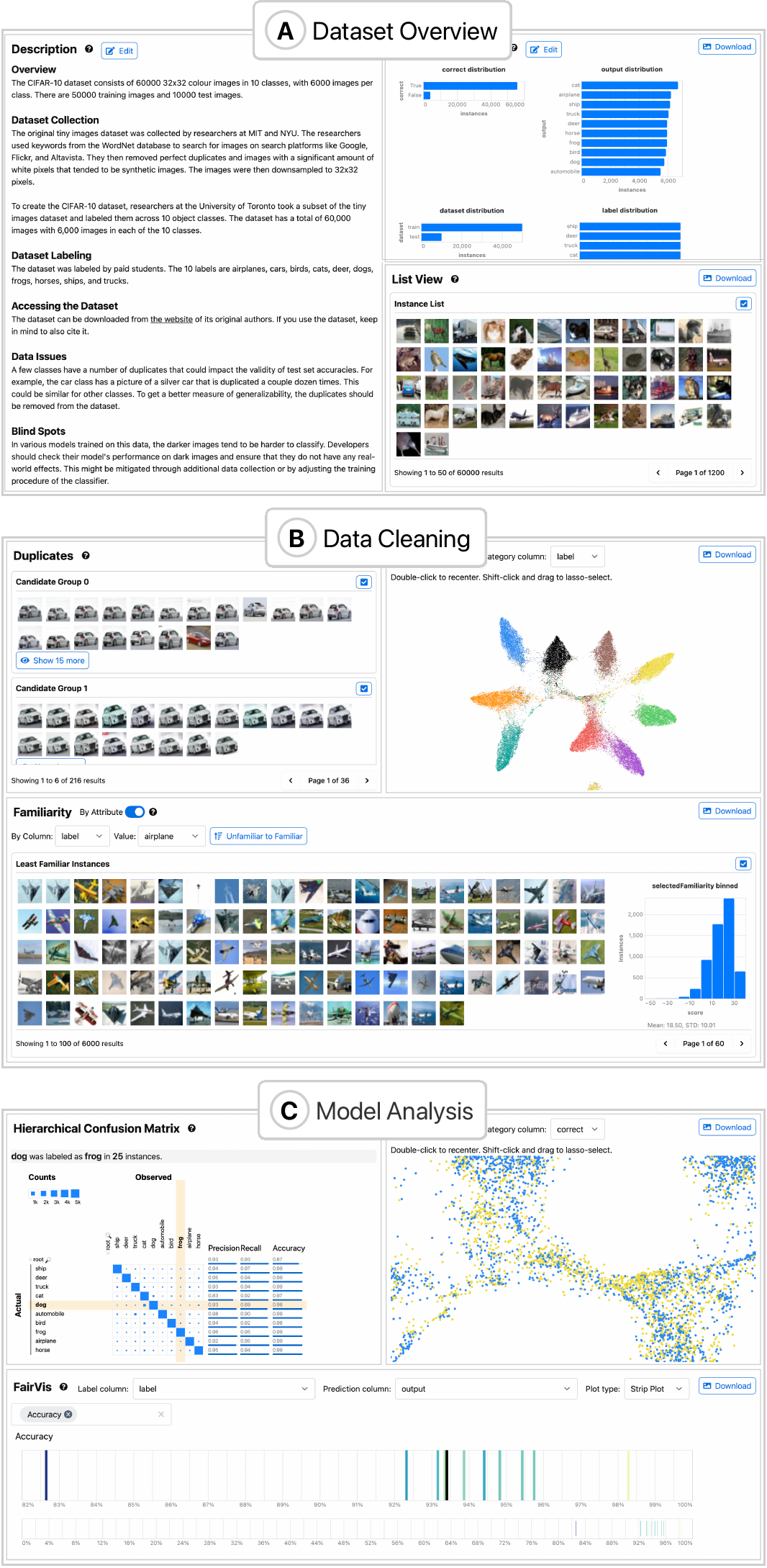}
    \caption{
    \system{} supports a diverse range of ML tasks. 
        Here we show examples of three distinct dashboards:
        (A) A dataset overview with a textual description of the data's origin, distribution plots, and example data instances.
        \creator{} can use this report to understand what a dataset contains and what tasks they can use it for.
        (B) A data validation dashboard to help \creator{}s track issues during data collection, such as duplicate or out-of-distribution instances.
        (C) A model analysis dashboard for exploring the performance of an ML system. 
        Users can find groups of incorrectly classified instances in the embedding and drill down into fairness metrics with respect to different data subgroups.
    }
    \Description{Three screenshots of example Symphony dashboards. The top left dashboard shows the dataset overview, the top right screenshot shows the data cleaning dashboard, and the bottom screenshot shows the model analysis dashboard.}
    \label{fig:types}
\end{figure}

The second wrapper we implemented is for standalone, web-based dashboards.
To support this, each \system{} component overrides an export function which is used by \system{} to transform selected visualization components from Python code into web-based UIs.
Components can be configured before export to be placed on different subpages and arranged within these pages to fit particular use cases, as shown in Figure~\ref{fig:types}.
These dashboards can be authored in programming environments and then exported as a statically hosted websites.
The wrapper for web-based UIs provides an HTML file which imports components as independent JavaScript (ES6) modules.
Since \system{} components are compiled to pure JavaScript files, the standalone dashboard does not need a dedicated backend and can be hosted on a static file server.

\newpage
\noindent\includegraphics[width=\linewidth]{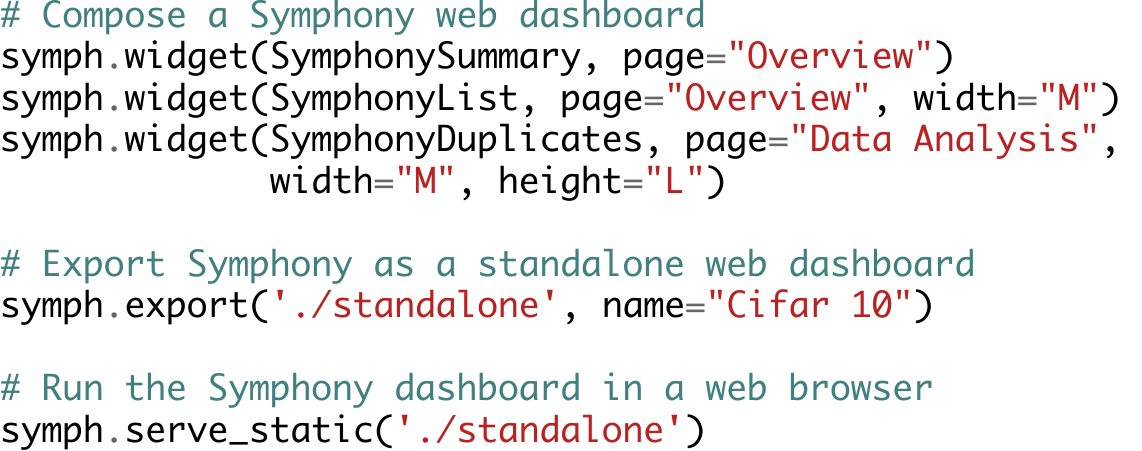}

New wrappers can be written to include \system{} components in other platforms.
For example, we began to explore how we can enable users without programming experience to create \system{} UIs using a drag-and-drop dashboard builder.
We have also experimented with integrating \system{} components in other interactive programming environments like Streamlit~\cite{Streamlit} or Glinda~\cite{DeLine2021a}.

\subsection{Interactive Exploration Tools}\label{sec:exploration}

The final key feature of \system{} is a set of tools for interacting with and exploring data.
Each component has the same interaction tools, and changes are reactively synchronized between components both in Jupyter notebooks and in web-based UIs.
For the web-based UI, state changes are also saved in the URL, allowing \consumer{}s to share specific findings.
\system{}'s interaction tools were derived both from common interactions described by participants in the formative study and findings from visualization research~\cite{Amar2005, Kang2007}.
We included a subset of tools that we found to be important for the specific components we implemented.
These tools include data filtering, grouping, and instance selection.
Additional interaction tools can be added to \system{} by updating the main \system{} package and platform wrappers with the new tool, which is then available on different platforms and synchronized across components.
New interaction tools can then be accessed and modified by individual \system{} components.

Users have three ways of using \system{}'s interaction tools: through a UI toolbar, \system{} components themselves, or code.
The UI toolbar (Figure~\ref{fig:component}, right) is available both in interactive programming environments (Figure~\ref{fig:notebook-and-overview}, left) and the web-based dashboards (Figure\ref{fig:notebook-and-overview}, right).
We implemented this toolbar as another \system{} component, which is shown alongside each component in Jupyter notebooks for convenient access, and displayed as a consistent sidebar for the web-based dashboard.
Apart from the UI toolbar, components not only have direct access to the global \system{} state but can also modify it based on user interaction.
For example, individual data samples can be selected from whichever component they are viewed in.
Thus, component developers can add custom controls to manipulate \system{}'s state.
Lastly, \creator{}s may want to make more complex data transformations that cannot be mapped to UI components.
For such use cases, \system{}'s state can also be directly be manipulated within Python.
Whether in a notebook or Python script, users can set and retrieve any of the state variables.
In Jupyter notebooks, this allows for fluid interactions between UI and code in the style of \citet{kery2020mage}.
Additionally, \system{}'s state can be extracted from the web-based UI and loaded into Python-based notebooks, making findings from shared \system{} dashboards available to the \creator{}s in code-based environments.

\vspace*{5pt}
\noindent\includegraphics[width=\linewidth]{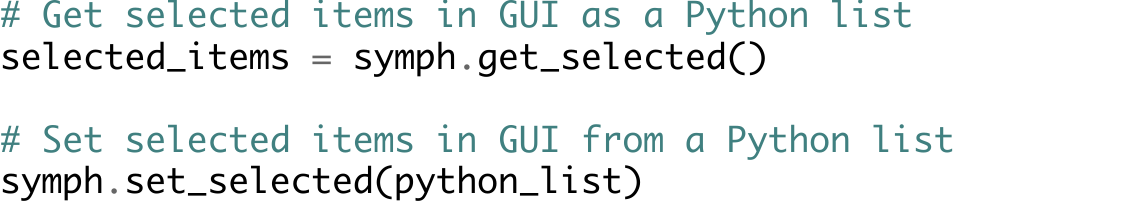}

\section{Participatory Design Sessions}

With the initial \system{} framework, we conducted a series of participatory design sessions to understand the specific needs of ML teams and design and develop an initial set of \system{} components.
We conducted 10 sessions where each session had between 1 and 7 people, with a total of 31 people across all sessions.
We recruited and contacted teams via internal mailing lists, and the sessions lasted between 30 minutes and an hour.
The first half of each session consisted of a demonstration of a \system{} prototype based on a mock dataset.
In the second half of each session, we asked participants to reflect on and describe their own work and asked them about what additional features would be necessary to integrate \system{} into their workflows.

\subsection{Expanding \system{}'s Technical Capabilities}

From these participatory design sessions, we extracted a set of additional needs and wants for \system{}.
Rather than the high-level goals presented in Section~\ref{sec:goalschallenges}, the findings from the participatory design sessions are more technical and tied to the implementation of \system{}.

While displaying images directly in computational notebook components was greatly appreciated by the participants working in computer vision, the teams working in different domains expressed interest in previewing and visualizing other data types.
To demonstrate \system{}'s ability to support other unstructured data types, we made the display of data sample modular and added audio data as an additional supported data type.
To visualize other types of data, a developer just has to implement a rendering function for the new data which all components can use.

\begin{figure*}
    \centering
    \includegraphics[width=\textwidth]{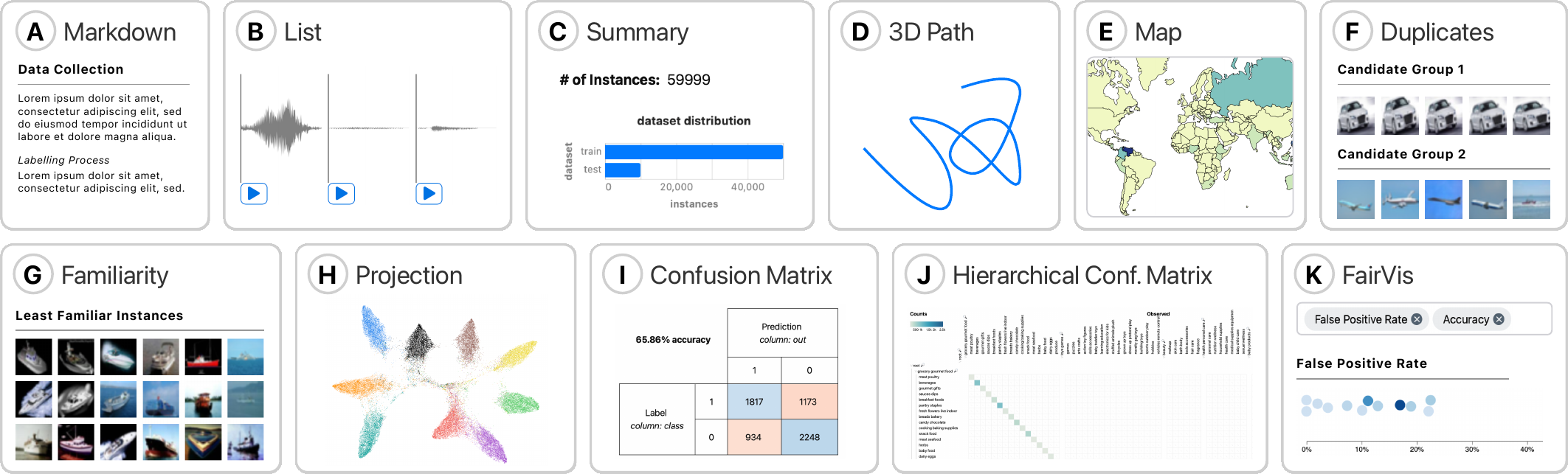}
    \caption{The \system{} components we implemented as a result of the participatory design sessions.
    (A) Markdown text for data and model details.
    (B) A paginated list of instances.
    (C) Distribution charts for metadata columns.
    (D) 3D path visualizations for sensor and inertial measurement unit (IMU) data.
    (E) Map visualizations for geographic data.
    (F) Potential duplicate instances.
    (G) Familiar and unfamiliar instances in a dataset.
    (H) 2D projection for model embeddings.
    (I) Binary confusion matrix.
    (J) Confusion matrix for hierarchical classification models.
    (K) Fairness analyses of intersectional subgroups.
    }
    \Description{A gallery of screenshots for the components available in Symphony.}
    \label{fig:components}
\end{figure*}

Some teams work with large models trained on big data, which originally exceeded \system{}'s ability to scale and led to long load times.
In response, we implemented pagination for all the components that display raw data.
Depending on the data type, the number of samples per page can be adjusted, allowing \system{} to scale to millions of data samples.
For even larger datasets, where a \creator{} wants to load and visualize hundreds of millions of data points, the browser memory becomes a limiting factor for holding the backing metadata table.
For these truly large datasets, we suggest users select representative subsets for detailed analysis; however, scaling beyond millions of data instances is described in Figure~\ref{sec:future-work}.

Interactive exploration is a powerful analysis technique when developing ML systems.
However, for ML projects that contain many datasets, compounded when data or models are rapidly changing, participants expressed interest in automatically generating shareable dashboards and reports to support streaming data and automatic model retraining.
Apart from providing \system{} as an authoring tool in computational notebooks, \creator{}s can also write Python scripts that consume ML data and model outputs, assemble a selection of components, and create and export a standalone \system{} web UI.

\subsection{Implemented \system{} Components for Data and Model Analysis}

Informed by the feedback and needs expressed in the participatory design sessions, we implemented an initial set of 11 components shown in Figure~\ref{fig:components}.
These initial components cover various data and model analysis tasks, from finding potential duplicates in a dataset to auditing models for biases.
We created all components using the component cookiecutter template described in Figure~\ref{sec:components}.

The first set of components created cover \textit{overview descriptions and summaries of an ML dataset}.
The markdown component (A) lets \system{} replicate existing documentation methods like Datasheets and Model Cards by writing rich text content. 
Users can follow existing guidelines to document essential information about a dataset or model often overlooked or not described. 
The list component (B) shows a paginated list of data instances, with support for a variety of data types like images and audio.
Multiple \creator{}s requested this feature, since they currently use file explorers outside of a notebook or one-off functions to look at individual instances.
Distribution charts and counts in the summary component (C) provide a high-level overview of data and can help detect potential biases or skews in a dataset.
Lastly, we developed two additional components, a 3D path component (D) and map component (E), for exploring specific data types like health sensor data and geographic distributions.

We also implemented a set of components for more \textit{complex analysis of unstructured datasets} that were important to multiple teams.
We first compute a model embedding from a deep learning model on the provided data instances, from which different metrics are calculated.
For the first of these components we use a nearest neighbors algorithm based on cosine distance in embedding space to find potential groups of duplicate instances (F), which could impact training performance or the validity of test set accuracy.
In the next component, we fit a Mixture of Gaussians model on the embeddings to calculate a familiarity score for each data point.
We find the most and least familiar instances in a dataset (G) by sorting by familiarity score. 
Instances with low familiarity scores can be outliers or mislabeled instances, while high familiarity instances can show over-represented types of data. 
Finally, there is a 2D projection embedding (H) that shows a dimensionality reduced representation of the embeddings.
The embedding can be used to find various interesting data and model patterns and is especially useful when used to explore insights found in other components.

The last set of components we implemented focus on \textit{analyzing and debugging ML models}.
The classic confusion matrix component (I) is important for initial debugging of classification models.
Other classification tasks that use data with hierarchical or multi-label data can be explored using a hierarchical confusion matrix component (J).
We primarily implemented this component for a team in the participatory design sessions that was working on hierarchical classification models.
Lastly, we built a set of visualizations for analyzing model performance across intersectional subgroups (G) based on a system by \citet{Cabrera2019}.
The visualization can help users audit their models for biases, something which multiple product teams were interested in.

We used three different methods for implementing the above \system{} components.
\system{} components are Svelte and JavaScript (JS) files, so authors can create new visualizations with their preferred front-end libraries.
For components without existing libraries, we used JavaScript in combination with visualization packages such as Vega and D3.
\system{} can also use off-the-shelf JS libraries, for example, we used REGL Scatterplot~\cite{reglplot}, a WebGL library, to create the projection component.
Lastly, since \system{} components are made with Svelte, we can also directly use Svelte components, which is what we did with the Hierarchical Confusion Matrix.
These different strategies for creating components provide the flexibility to implement custom visualizations while also allowing developers to use off-the-shelf libraries and visualizations.

\section{Case Studies on Deployed ML Systems}

Lastly, we evaluated \system{} with \creator{}s and \consumer{}s working on real-world ML products.
We worked with three ML teams at Apple, drawn from the participatory design sessions, to integrate \system{} with their data and ML pipelines.
The teams focus on different machine learning tasks, namely dataset creation and labeling, accessibility research, and ML education.
To understand the affordances and limitations of \system{}, we conducted think-aloud studies lasting 60 minutes where a member of each team used \system{} to explore a Jupyter notebook and create a web-based dashboard for their data and model.
While field studies like ours excel at capturing how participants actually work, this data has to be collected opportunistically.
We believe these case studies capture the target audience of \system{}, cross-functional teams working on modern ML models trained on unstructured data, but may have some insights specific to organizational workflows.

Before the study, we sent a member of each team, the main participant, a Jupyter notebook that imported their data and displayed a set of \system{} components applicable to their domain and task.
The study was split into three main sections.
For the first third of the study, we asked the team to think aloud while the main participant used the notebook and \system{} components to explore the data and model freely.
In the second part of the study, we asked the main participant to export the \system{} components (using a command in the notebook) to a standalone dashboard and continue exploring in the exported web UI.
For the final part of the study, we asked the team for feedback on \system{} and discussed what types of use cases or limitations they found.

\subsection{Case Study I: Validating and Sharing Data Patterns on a Dataset Creation Team}

For the first case study, we worked with a team that assembles and labels large machine learning datasets.
Their datasets are composed of labeled images and videos which they publish to an internal data repository.
The team was interested in using \system{} in two ways, first, using it during dataset creation to detect errors in the data and labels, and second, as a reporting tool to give consumers of the dataset details about the data.
Given these requirements, we loaded \system{} with the list (Figure~\ref{fig:components} (B)), summary (Figure~\ref{fig:components} (C)),  duplicates  (Figure~\ref{fig:components} (F)),  familiarity  (Figure~\ref{fig:components} (G)),  projection  (Figure~\ref{fig:components} (H)), and map  (Figure~\ref{fig:components} (E)) components.

The main participant started in the notebook and used multiple components and interaction tools in concert to spot unexpected patterns in their data. 
They made extensive use of \system{}'s toolbar to combine filters and select subsets of data in which they were interested.
When using the notebook, they commented that \feedback{there are a lot of neat things here, first, the filter carried over, and it is so cool to see the data samples and metadata within the notebook.}
The synchronized, reactive state let them validate insights from the filtered summary charts with the actual raw instance previews in the list view.
Next, the main participant moved on to the duplicates and familiarity components, where they found a couple of labeling errors that they suspected existed in their dataset but had not been able to validate previously.
After transitioning to the standalone dashboard, the first component they looked at was the projection visualization.
They used the projection to find a closely clustered group of instances where a few highlighted points that the model had misclassified.
In the standalone dashboard, they also dubbed the map visualization \feedback{very useful}, especially when sharing reports of their data collection efforts with managers or policymakers.

Overall, the team found \feedback{a lot of value here} when using \system{}.
They mentioned that the workflow they would most prefer would be automatically generating shareable reports for every dataset they published: \feedback{programmatic generation and live visualizations are awesome, being able to pop these charts into all our READMEs would be amazing.}
They saw the standalone dashboard that they created with \system{} as a \feedback{great starting point} for analyzing their datasets, and that they could see people use the notebooks for more detailed analysis: \feedback{if people want to drill down more, and get exact specific access, summon the notebook.}
Being able to create different interfaces with subsets of visualization components was important for them as well, as different audiences have different needs and they \feedback{do not want customers to do the data cleanup} for them.

The team also identified usability issues and limitations in \system{}.
When initially using the projection component, the main participant was not sure what it showed and thought that \feedback{this component would need some introduction, as it has complex controls.}
They also requested additional components, such as heatmaps and other 2D graphs, to do a more detailed analysis of distributions.
Lastly, the main limitation for directly using \system{} was not being able to attach the raw data files to a \system{} interface as their data samples are often not hosted and too large to duplicate.

\subsection{Case Study II: Debugging Training Data on an Accessibility Team}

In the second case study, we worked with a team that uses ML to make software applications more accessible.
They have a large dataset of icon screenshots for which we assembled a similar set of components to the dataset creation team.
We included the summary (Figure~\ref{fig:components} (F)), duplicates (Figure~\ref{fig:components} (A)), familiarity (Figure~\ref{fig:components} (B), and projection (Figure~\ref{fig:components} (H)) components .

When exploring the notebook, the participant found the duplicates, familiarity, and scatterplot components to be the most interesting.
Since they use an automated approach to collect their data, the participant assumed that there were likely duplicates in the dataset but had protocols to ensure they would not be across the training and testing set.
Using the duplicates component, they confirmed that a significant number of icons were duplicates, but when they used the grouping interaction to split the data by testing and training they found that a significant number of instances were duplicated across the two datasets.
The combination of the duplicates visualization and grouping interaction tool helped them discover that they \feedback{were cheating learning on samples we test for.}
The participant identified the problematic duplicates and selected them in the notebook to remove from the test set with a Python command later.
Next, the participant explored the familiarity component and found a large number of similar grey icons, based on which they wondered if \feedback{the model might overfit on these samples.}
Finally, using the projection visualization, they found a dispersed cluster of instances with different labels.
When they selected the group, they found that the instances were all PNG images in the test set, while the training set only contained JPEG images.
The participant then mentioned they \feedback{want to test their model specifically on PNG images to assess how the model generalized.}

Overall, the participant mentioned that they would \feedback{want to try and use this to share insights within the team.}
Additionally, they found the notebook-based visualizations personally useful to \feedback{look into the data,} which they had previously done manually using a file explorer outside of the notebook.
They mentioned that they would likely use a computational notebook to explore data, and only use the standalone dashboards to share insights or when they wanted more visualization space.
The main feature the team wanted was to combine data and model findings to understand the impact of data changes: \feedback{it would be super helpful to also add models and combine model analysis with existing components.}
While this analysis is possible with existing model analysis components, future components could specifically combine data and model information.

\subsection{Case Study III: Promoting Data Exploration for ML Novices on an Education Team}

For the final case study, we collaborated with a team focused on ML education.
They teach courses about ML principles and techniques to engineers, and also teach their audience about data and model analysis tools.
They sent us a list of datasets they commonly explore with students from which we selected two representative datasets, one audio dataset for data analysis and one image dataset for model analysis.
For the audio dataset we used the same components as in the previous evaluation.
To support model analysis for the image dataset, we used the summary (Figure~\ref{fig:components} (F)), hierarchical confusion matrix (Figure~\ref{fig:components} (D)), FairVis (Figure~\ref{fig:components} (K)), and projection (Figure~\ref{fig:components} (H)) components.

The team was interested in how they could use separate components in concert.
They used the cross-filtering and grouping heavily to combine, for example, the projection visualization with the summary component to spot misclassified samples.
They also used the confusion matrix visualization in combination with our filtering tool.
For example, they filtered out the correctly classified data samples from the metadata table to highlight misclassifications and described the resulting confusion matrix as \feedback{a fantastic graphic.}
They were also intrigued by being able to display a list of data samples in notebooks or a standalone dashboard, as they \feedback{constantly tell [their] students to look at a lot of examples} but are currently limited to seeing one or two instance at a time and \feedback{just graph things using matplotlib or pillow.}

Overall, the team found \system{} to be a valuable tool for ML tasks and thought it could play a part in one of their lessons, as \feedback{promoting looking at data is extremely important.}
They remarked that \feedback{in Python, its very easy to ignore the data, anything you can do to bring the data to the forefront is great.}
Thus, they wanted to use \system{} during their courses in multiple ways, namely using the \feedback{notebook for generating interfaces, then exporting them to teach a group such that they can open the website and everyone can explore on their own or follow my instructions.}
This way, they hoped that \feedback{[students] can play with it and experiment talk about how to communicate results for ML models.}
They also particularly liked the option to assemble visualizations, for example when their students learn how to \feedback{communicate findings to executives} and \feedback{graphing the relevant, and hiding the irrelevant.}

As for limitations, they wanted to be able to unlink the state of different components to experiment with them independently.
They also mentioned that they would like to load more data types than just the currently implemented audio, images, and tabular data, namely text data.
While this is not possible right now, \system{} could be extended to more data types by augmenting the data sample adapter we provide.

\section{Limitations and Future Work}
\label{sec:future-work}
In both the pilot studies and case studies, we found ways in which \system{} could be further improved.

\paragraph{Authoring components}
\system{} components are written using JavaScript code and web-based visualization libraries.
Programming these visualizations requires expertise in web development and visualization, which limits who can create new components. 
Future work could explore ways to lower the barrier to authoring new visualization components.
Potential strategies to make component creation more accessible include using grammars for interactive graphics, such as Vega~\cite{satyanarayan2015reactive}, or UI-based visualization builders like Tableau~\cite{Tableau}.
Additional research would be needed to make these tools more expressive for unstructured data and ML models.

\paragraph{Scaling past millions of data points}
\system{} currently loads the backing metadata table used for \system{} into web browser memory.
This scales to tens of millions of data points, which, while sufficient for many modern ML tasks, does not cover all domains.
In our design sessions, we spoke to teams with terabyte-scale metadata tables that do not fit in browser memory.
Future work could explore ways to support this scale while still providing direct interactivity with the underlying data and models.
Using an external API or backend for data processing combined with more efficient data queries could support massive data but would limit where the web-based UI could be used.

\paragraph{Beyond conventional data science platforms}
In this work, we implemented \system{} wrappers for computational notebooks, programming environments, and web-based dashboards.
While these platforms cover a significant portion of where ML work happens, future work could explore how \system{} could be incorporated into other platforms, especially those which are currently isolated from data science work.
For example, \system{} interfaces could be included in messaging services, documents, presentations, or issue trackers to further bring the benefits of \system{} to more people.
New design studies could be conducted to understand how users in common communication platforms like instant messaging would benefit from and use \system{} components.

\paragraph{Guided usage of Symphony}
\creator{}s can use \system{} for a wide array of ML analyses, from dataset debugging to auditing models for bias. 
This gamut of uses stands in contrast to more \textit{prescriptive} approaches like Datsheets~\cite{gebru2018datasheets}, Model Cards~\cite{mitchell2019model}, and checklists~\cite{Madaio2020} which define an ordered list of what an ML interface should show.
While \creator{}s can use \system{} for more types of analyses, it does not provide any guidance to users about which components might be the most adequate or useful for a given task.
Future work could look at combining \system{}s open-ended, exploratory approach with more prescriptive guidance.

\paragraph{Scope of case study findings}
Lastly, our case studies were conducted with \creator{}s at a single institution that works on large ML models trained on unstructured data, often using notebooks and visualization dashboards.
While we believe these tools and ML development practices exist widely in industry and academia, we recognize some of our findings may not generalize to other organizations or types of users such as machine learning enthusiasts, hobbyists, or small teams.
Further studies could explore \system{}'s affordances and drawbacks in these distinct settings.

\section{Discussion}

\system{} provides a common substrate for ML interfaces that enables both exploratory analysis and sharable ML interfaces.
By meeting different users where they work, \system{} empowers each member of an ML team to have direct access and knowledge of the data and models powering an AI product.

While the case studies described scenarios where \creator{}s work in programming environments and then transition to web-based UIs, we also observed in our studies that \creator{}s can benefit from going the opposite direction: transitioning from a web-based UI back to a programming environment.
When a user finds an interesting insight in a standalone \system{} dashboard, they can copy their findings to the programming environments along with state variables like filters and groups.
Existing analysis tooling often suffers from an ``expressiveness cliff'', where only a fixed set of visualizations and data manipulations is available.
\system{} allows users to return to programming environments where they have more flexible analysis tools.

\creator{}s' desire to use \system{} for exploration could also encourage them to share their insights more frequently.
If \creator{}s are using a set of \system{} components for exploratory analysis in a notebook, no additional work is needed for them to export it as a standalone, shareable UI.
Participants mentioned the ability to programmatically combine components as a major benefit, allowing them to go from exploration to an interactive, web-based UI without using a different tool.
Additionally, \system{} interfaces can be redeployed continuously whenever the data and model are updated, supporting ML tasks with streaming data or automatic model retraining.

By integrating with existing data science platforms, \system{} could also encourage broader use of task-specific ML visualizations.
ML visualization systems are often implemented as one-off web dashboards~\cite{Cabrera2019,hohman2019summit,Ahn2019,Wu2019,Cashman2019,Krause2016} that require users to wrangle and export their data into systems separate from where they do ML development.
\system{} includes task-specific visualization components directly in data since platforms like Jupyter notebooks, and the components can consume data from standard data APIs like Pandas Data Frames.
In turn, implementing ML visualizations as independent components in a framework like \system{} could increase their use and longevity.

Beyond helping individuals understand ML systems, \system{} is intended to foster a shared organizational understanding~\cite{Cabrera2002} between stakeholders on an ML team.
\system{} interfaces act as \textit{boundary objects} for large, cross-functional ML teams. 
Boundary objects are artifacts that are \textit{``both plastic enough to adapt to local needs and the constraints of the several parties employing them, yet robust
enough to maintain a common identity across sites''}~\cite{star1989institutional}.
\system{} can serve as a boundary object for ML teams, providing interfaces that adapt to the different needs of \consumer{}s.
At the same time, \textit{``The creation and management of boundary objects is a key process in developing and maintaining coherence across intersecting social worlds''}~\cite{star1989institutional}.
\system{} aids in this creation and management process, bridging the gap between the intersecting worlds of different ML stakeholders such as engineers, designers, and product managers.

\section{Conclusion}

In this work, we designed and implemented \system{}, a framework for composing interactive ML interfaces with data-driven, task-specific visualization components.
\system{}'s visualizations helped ML teams find important issues such as data duplicates and model blind spots.
Additionally, We found that by providing ML interfaces in the data science platforms where \creator{}s work, \system{} can encourage \creator{}s to \emph{want} to use and share insights.
With data-driven components that diverse \consumer{}s across an ML team can use, \system{} fosters a culture of shared ML understanding and encourages the creation of accurate, responsible, and robust AI products.

\begin{acks}
We thank our colleagues at Apple for their time and effort integrating our research with their work.
We especially thank Kayur Patel for his guidance and Mary Beth Kery for her generosity reviewing early drafts of this work.
\end{acks}

\bibliographystyle{ACM-Reference-Format}
\bibliography{main}

\end{document}